\documentclass[onecolumn,preprintnumbers,amsmath,amssymb,superscriptaddress]{revtex4-2}
\usepackage{epsf}
\usepackage{graphicx}
\usepackage{sidecap}
\usepackage{array}
\usepackage{amsmath}
\usepackage[utf8]{inputenc}
\usepackage{amssymb}
\usepackage{dcolumn}
\usepackage{epstopdf}
\usepackage{bm}

\usepackage{color}

\def \beq {\begin{equation}}
\def \eeq {\end{equation}}
\begin{document}
\title{{Anisotropically large anomalous and topological Hall effect in a kagome magnet  }}
\author{Gyanendra~Dhakal*} 
 \affiliation {Department of Physics, University of Central Florida, Orlando, Florida 32816, USA}
\author{Fairoja Cheenicode Kabeer*} 
\affiliation{Department of Physics and Astronomy, Uppsala University, P. O. Box 516, S-75120 Uppsala, Sweden}
\author{Arjun K. Pathak*} 
\affiliation{Department of Physics, SUNY Buffalo State, Buffalo, New York 14222, USA}
\author{Firoza~Kabir*} 
\affiliation {Department of Physics, University of Central Florida, Orlando, Florida 32816, USA}
\author{Narayan Poudel} \affiliation{Idaho National Laboratory, Idaho Falls, ID 83402, USA}
\author{Randall Filippone} \affiliation{Department of Physics, SUNY Buffalo State, Buffalo, New York 14222, USA}
\author{Jacob Casey}\affiliation{Department of Physics, SUNY Buffalo State, Buffalo, New York 14222, USA}
\author{Anup Pradhan Sakhya}\affiliation {Department of Physics, University of Central Florida, Orlando, Florida 32816, USA}
\author{Sabin Regmi}\affiliation {Department of Physics, University of Central Florida, Orlando, Florida 32816, USA}
\author{Christopher Sims}\affiliation {Department of Physics, University of Central Florida, Orlando, Florida 32816, USA}
\author{Klauss Dimitri}\affiliation {Department of Physics, University of Central Florida, Orlando, Florida 32816, USA}
\author{Pietro Manfrinetti} \affiliation{Department of Chemistry, University of Genova, 16146 Genova, Italy} \affiliation{Institute SPIN-CNR, 16152 Genova, Italy}
\author{Krzysztof Gofryk} \affiliation{Idaho National Laboratory, Idaho Falls, ID 83402, USA}
\author{Peter M. Oppeneer}\affiliation{Department of Physics and Astronomy, Uppsala University, P. O. Box 516, S-75120 Uppsala, Sweden}
\author{Madhab Neupane $\dagger$}
\affiliation {Department of Physics, University of Central Florida, Orlando, Florida 32816, USA}
 
\begin{abstract}{Recently, kagome materials have become an engrossing platform to study the interplay among symmetry, magnetism, topology, and  electron correlation. The latest works on $R$Mn$_6$Sn$_6$ ($R$ = rare earth metal) compounds have illustrated that this family could be  intriguing  to investigate various physical phenomena due to large spin-orbit coupling and strong magnetic ordering. However, combined transport and spectroscopic studies in $R$Mn$_6$Sn$_6$ materials  are still limited.  Here, we report magnetic,  magneto-transport, and angle-resolved photoemission spectroscopy measurements of a kagome magnet ErMn$_6$Sn$_6$ that undergoes  antiferromagnetic ($T_N$ = 345 K) to  ferrimagnetic ($T_{C}$ = 68 K) phase transitions {in the presence of field}. We observe  large anomalous  and topological Hall effects serving as transport signatures of the nontrivial Berry curvature.  The isothermal magnetization exhibits strong anisotropic  nature and the topological Hall effect of the compound  depends on the critical field of metamagnetic transition. Our spectroscopic results complemented by theoretical calculations show the multi-orbital kagome fermiology. This work provides new insight into  the tunability and interplay of topology and magnetism in a  kagome magnet. }
\end{abstract}
\pacs{}
\maketitle
\noindent
\noindent

 The interplay among  magnetism, correlation, and topology has captured enormous attention currently owing to exotic transport and electronic behaviors \cite{keimer2017physics, sachdev2018topological, he2018topological, hosen2018discovery,   PhysRevB.81.245209}. The kagome lattice, a two-dimensional network of a triangular Bravais lattice sharing the corners, provides an opportunity to study diverse quantum magnetic phases \cite{sachdev2018topological, han2012fractionalized, RN215, yin2019negative, RN227, PhysRevLett.112.017205}. 
  Due to the unusual lattice geometry and breaking of time-reversal symmetry, kagome magnets can support Dirac fermions  \cite{RN216, han2012fractionalized}, intrinsic Chern quantum phases \cite{PhysRevLett.106.236802, PhysRevLett.115.186802}, and spin liquid phases \cite{RN218} leading to a sudden surge of  research interest in these materials.  Kagome lattices possessing 3\textit{d} transition metals such as Fe$_3$Sn$_2$ and Co$_3$Sn$_2$S$_2$ have bolstered the  research interests as they exhibit large Berry curvature fields and giant magnetization driven electron nematicity \cite{RN219, RN220, RN215, RN222, PhysRevLett.121.096401, RN223, PhysRevB.94.075135,  RN226, Nayake1501870}. Kagome systems with spin-orbit coupling (SOC) and out-of-plane ferromagnetic order  fulfill   conditions for Chern gapped Dirac fermion, that can give rise to quantum anomalous Hall effect \cite{PhysRevLett.106.236802, PhysRevLett.115.186802}. It is worth noting that  binary kagome systems containing transition metals usually lack strong out-of-plane magnetization, therefore recent works are focused on the materials with strong SOC and out-of-plane magnetization which can provide an epitome of kagome materials in which one can pursue the topologically gapped Dirac fermion \cite{ma2021anomalous, PhysRevLett.115.186802, RN227,ma2020rare}. 
 One  such family is rare earth ($R$) based $R$Mn$_6$Sn$_6$ kagome magnet, where presence of  SOC and strong magnetic order serve as an intriguing quantum key to realize near-ideal quantum limit Chern magnet \cite{ma2020rare, RN227}.
 
 These rare earth 166 compounds usually reveal magnetic ordering at room temperature, with
antiferromagnetic (AFM) ordering for non-magnetic $R$ ions and ferrimagnetic (FIM) ordering for
4\textit{f}- local-moment-carrying $R$ ions \cite{clatterbuck1999magnetic, malaman1999magnetic}. The  compounds with $R$ = Gd-Ho are ferrimagnetic below the Curie temperature, whereas the ones with $R$= Sc, Y, Tm
and Lu exhibit antiferromagnetic or helimagnetic behaviors. 
 The magnetic structure of
$R$Mn$_6$Sn$_6$ compounds consists of two different subsystems: the $R$ and Mn subsystems. The
interlayer Mn moments through the Mn-Sn1-Sn2-Sn1-Mn planes along the $c$-axis are always {ferromagnetic whereas Mn-$R$-Mn and Mn-Sn-Mn magnetic moments interactions depend on the nature of
the rare earth elements \cite{yazdi2012magnetoelastic}}. Interestingly, these compounds show Shubnikov-de Haas quantum oscillations below 14 T field identifying themselves as distinct type of kagome magnets \cite{ma2020rare}. 
Despite a few works in transport measurements \cite{ma2020rare, RN227, Ghimireeabe2680,  PhysRevLett.106.236802, PhysRevB.103.014416, mielke2021intriguing}  comprehensive magnetic, transport and spectroscopic measurements of the electronic structures of these materials are still limited, therefore,  understanding of the potentially interesting obscured quantum phenomena are in demand. Among these materials,  ErMn$_6$Sn$_6$ shows a complex magnetic behaviour displaying several transitions. {At low temperature (T$<$65 K) and under low or zero magnetic field, it exhibits AFM (namely AFM1) ordering due to antiparallel alignment of ferromagnetic sublattices of Er and Mn, yet even  low field can change AFM to FIM state. In the intermediate temperature range 65 K$\leq$ T $\leq$ 350 K, it contains an AFM Mn sublattice and a paramagnetic (PM) Er sublattice resulting AFM state (namely AFM2) which changes to FIM under the influence of strong field. Beyond, 350 K it shows PM due to PM sublattices of Er and Mn \cite{malaman1999magnetic, clatterbuck1999magnetic, yao2004unusual, suga2006high}. } 
Therefore, ErMn$_6$Sn$_6$  could provide a unique system to study the evolution of transport and electronic structures with magnetic regimes.\\

In this article, we report magnetic, magneto-transport, and electronic structure measurements of   ErMn$_6$Sn$_6$ in two magnetic regimes.  Transport measurements show anisotropically large anomalous  Hall effect (AHE) and topological Hall effect (THE) as an indication of non-trivial topology linked with the kagome material. Our angle-resolved photoemission spectroscopic (ARPES) measurements supported by first-principles calculations demonstrate the presence of multi-orbital kagome bands in the vicinity of the Fermi level with the occurrence of a Dirac-like dispersion at the K point. Our results provide a novel platform to investigate the conundrum  of interplay between magnetism and topology in  kagome materials. \\
\begin{figure}
	\centering
	\includegraphics[width= \textwidth]{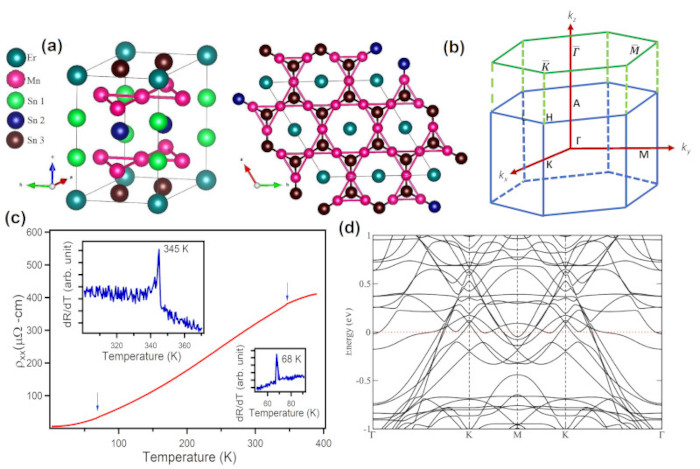}
	\caption{Crystal structure and sample characterization of  ErMn$_6$Sn$_6$.
(\textbf{a}) {Crystal structure of ErMn$_6$Sn$_6$. Right panel shows the top view of crystal structure forming the kagome lattice.} (\textbf{b}) {Bulk 3D Brillouin zone (blue) along with the projected [001] surface Brillouin zone (green), where high symmetry points are labeled.} (\textbf{c}) Electrical resistivity measured as a function of temperature in zero field. The insets on the top left corner  and the bottom right corner show the magnetic transitions.   (\textbf{d})  First-principles calculations of the bulk band of ErMn$_6$Sn$_6$ with considering spin-orbit coupling (SOC). } 
\end{figure}
High-quality single crystals of ErMn$_6$Sn$_6$ were synthesized using the Flux method as explained in the supplemental material \cite{SI}. Resistivity and Hall measurements were performed in a Physical Property Measurement System (PPMS) following the conventional 4-probe method. ARPES measurements were performed  at the 5-2 endstation in 
Stanford Synchrotron Radiation Lightsource (SSRL) beamline using a DA30L analyser.  The first-principles electronic structure calculations are carried out within the density-functional theory (DFT) formalism  \cite{Hohe64M,Kohn65M} as implemented in the Vienna \textit{ab initio} simulation package (VASP) \cite{Kres96M,Kres96.1M}.  For details, see Sec. 1.4 of the SM \cite{SI}  and Refs.\cite{Hohe64M, Kohn65M, Kres96M,Kres96.1M, Bloo94M, Perd96M, Duda98M, Monk76M, Lope85M, Most14M, Wu18M} therein for details.\\

 ErMn$_{6}$Sn$_{6}$ crystallizes in the hexagonal MgFe$_6$Ge$_6$-type structure with space group P6/mmm
(No. 191)  as shown in Fig. 1(a) with Er at 1(a) (0, 0, 0), Mn at 6(i) (1/2, 0, z$\sim$0.249), Sn at 2(c) (1/3, 2/3, 0), 2(d) (1/3, 2/3, 1/2) and 2(e) (0, 0, z$\sim$0.34) \cite{malaman1999magnetic}. 
 The Er and Sn3 atoms lie in the same plane and
Mn-Sn1-Sn2-Sn1-Mn atoms are stacked along the c-axis alternatingly. The Mn atoms form
two kagome layers and the Sn2 and Sn3 atoms form a hexagonal structure. The hexagonal
structure formed by Sn3 atoms can be clearly seen from the right panel of Fig. 1(a) when viewed from the
c-axis. Since Sn2 is below the Sn3 layers, it cannot be seen from this top view. The Er
atoms lie at the centre of the hexagons formed by the Sn3 atoms. The Sn1 atoms 
lie at the centre of the hexagons of the Mn layers but they are below or above those Mn
layers. The magnetic configuration is characterized by the ferromagnetic (001) Mn planes
\cite{yao2004unusual}. The Mn planes and the in-plane nearest neighbor Mn-Mn bonds d$_{Mn-Mn}$= 2.75 $\AA$
are
crystallographically equivalent but the interplanar Mn-Mn bonds along the c-axis are different
d$_{Mn-Mn}$= 4.5 $\AA$. The complex magnetic phase transitions in ErMn$_6$Sn$_6$ originate from the
complicated interplay among the $R$-Mn, $R$-$R$ and Mn-Mn exchange interactions and the competing magnetocrystalline anisotropies of the two sublattices \cite{yazdi2012magnetoelastic}. In the antiferromagnetic
state, metamagnetic behavior was also observed \cite{malaman1999magnetic} and neutron diffraction studies have confirmed a high-temperature helimagnetic structure and a low-temperature
ferrimagnetic structure in this compound \cite{malaman1999magnetic}. In Fig. 1(b) the three-dimensional Brillouin zone is presented, which is projected onto the (001) plane forming a hexagonal pattern. Resistivity as a function of temperature,  shown in Fig. 1(c), indicates the metallic nature of the sample over the measured temperature range of 2 K - 400 K. The derivative of the resistance shows two distinct peaks: the inset on the bottom right corner represents the magnetic phase transition ($T_{C}$ = 68 K), and the inset on the top left corner represents another magnetic phase transition ($T_N$= 345 K). 
    The first-principles calculation with SOC at the FIM phase is  presented in Fig. 1(d), which shows that multiple bands  cross the Fermi level indicating  ErMn$_6$Sn$_{6}$ to be a  metal which is in accord with the transport measurements. Electron-like bulk bands are seen along the $\Gamma$-K direction. Two Dirac dispersions are seen at the K point, one Dirac cone appears above the Fermi level, whereas another Dirac cone lies below the Fermi level.   
     A flat band is seen $\sim$400 meV above the Fermi level. 
      The bulk-band calculations in the AFM-state show some  changes in the electronic structures, such as shifting of flat bands upwards (see SM for bulk-band calculations \cite{SI}).
  \begin{figure*}
	\centering
	\includegraphics[width=\textwidth]{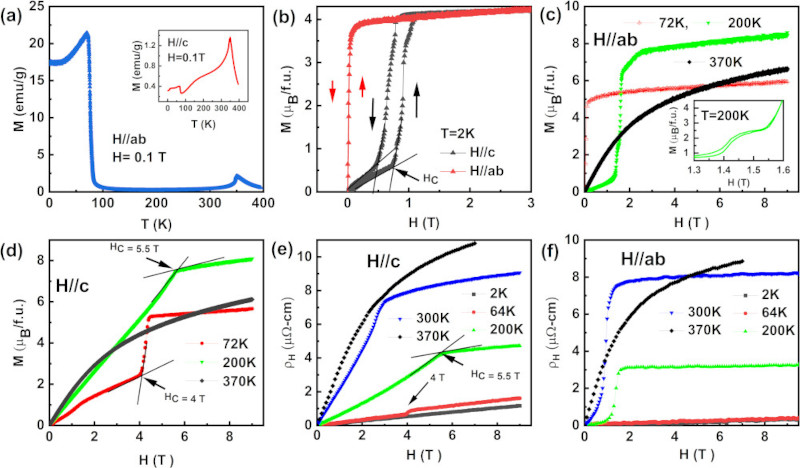}
	\caption{ Magnetic properties of ErMn$
	_6$Sn$_6$ single crystals measured along both $H//ab$ and H//c planes. (a) Magnetization as a function of temperature, M(T), measured along $H//ab$ at H = 0.1 T.  Inset in (a) shows M(T) measured along $H//c$ plane at H = 0.1 T.  (b) Magnetization as a function of magnetic field, M(H) measured along both $H//ab$, and $H//c$ plane at T = 2K. The M(H) was measured up to 9 T, only up to 3 T is shown for clarity. (c) and (d) Magnetization as a function of magnetic field measured along $H//ab$, and $H//c$, respectively at different temperatures. Inset in (c) shows selected part of M(H) to show the metamagnetic transition around 1.4 T  at T = 200 K. Hall resistivity of ErMn$_6$Sn$_6$ at different temperatures with (e) $H//c$, and (f) $H//ab$ direction. }
\end{figure*}
 
Magnetic properties as  functions of both temperature and magnetic field for ErMn$_6$Sn$_6$ were studied along $H//ab$ and $H//c$ planes by physical property measurement system (PPMS, Quantum Design) between temperatures 2 to 400 K and magnetic field up to 9 T.  The iso-field magnetization, M(T), measured along  $H//ab$  at H = 0.1 T is shown in Fig. 2(a).  The compound orders antiferromagnetically at T$_N$ = 345 K and the magnetization is almost zero between temperatures 300 and 100 K, suggesting the  AFM state. The isothermal magnetization curves, M(H), at a low magnetic field further support the AFM state in this temperature region (see  Fig. 2(c)). Upon further cooling, the exchange interaction between Er and Mn sublattices becomes strong resulting in the ferrimagnetic (FIM) behavior with T$_C$ = 68 K {at H = 0.1 T for $H//ab$. The isothermal magnetization curves, M(H), at intermediate temperatures at a low magnetic field, further support the AFM state (see Fig. 2(c)). } The application of the magnetic field shows a strong effect on the magnetic structure along the $H//c$. {The M(T) measured along $H//c$ shows a similar transition in $H//ab$, however, the nature of the magnetization curves are quite different and highly anisotropic (inset, Fig. 2(a)). The value of magnetization is negligible compared to that for in $H//ab$.  Comparing the M(H) for both $H//ab$ and $H//c$ (Fig. 2c, 2d), the transitions require a larger applied field for H$//$c.  At T = 2K and $H//ab$,}   magnetization initially increases almost linearly with field until H$\leq$0.7 T then a sharp increase in magnetization occurs at a critical magnetic field H$_{c}$ $\sim$0.7 T, indicating the field-induced metamagnetic transition (see Fig. 2(b)).  A strong change in magnetization occurs both in magnetizing and demagnetizing field with a relatively large magnetic hysteresis of 0.63 T, suggesting a field-induced first-order transition.  The metamagnetic transition indicates the field-induced transformation from AFM to FIM state in ErMn$_6$Sn$_6$. Unlike along $H//c$, the M(H) measured along the $H//ab$, however, does not show a metamagnetic transition at the corresponding temperature, suggesting easy magnetization direction along the $ab$-plane. {As reported by Clatterbuck and Gschneidner \cite{clatterbuck1999magnetic} the application of a magnetic field increases the low-temperature transition while decreasing the high-temperature transition, and the two transitions merge at H $\sim$ 1.75 T. 
}\\
\begin{figure}
	\centering
	\includegraphics[width= 0.9\textwidth]{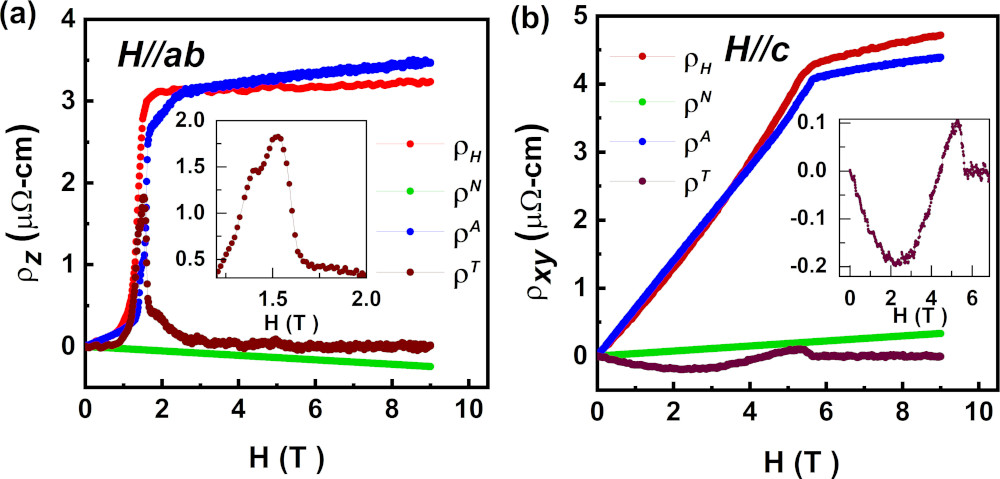}
	\caption{ {Realization of anomalous and topological Hall effect.} Hall resistivity and its three different components in (a) $H//ab$-plane, and  (b) in $H//c$-axis  at a temperature of  200 K.  Inset in (a) and (b) shows $\rho^T$ vs H for clarity along $H//ab$, and $H//c$, respectively    }
\end{figure}
 The saturation magnetic moments (M$_s$),  obtained by M vs. H$^{-1}$ graphs by extrapolating the curves to H$^{-1}$ = 0 are 4.7 $\mu_B$/f.u. and 4.3 $\mu_B$/f.u. along $H//ab$ and $H//c$, respectively, which shows the FIM structure at high field. The critical magnetic field to induce the AFM to FIM state for $H//c$ increases until T = 170 K and slowly and decreases with further increase in temperature (e.g. Hc $\sim$5.5 T for T = 200 K).  At T = 200 K and H = 9 T, the magnetic moment is 8.47 and 8.06 $\mu_B$/f.u. along the $H//ab$ and $H//c$, respectively which is twice the value of Ms obtained at T = 2 K. 
 
In order to investigate the signature of the anomalous quantum Hall effect, we present the transverse resistivity response as a function of the magnetic field applied parallel to the $c$-axis and $ab$-plane as depicted in Figs. 2(e) and 2(f), respectively. The Hall resistivity follows  different trends in the different magnetic regimes. In the ferrimagnetic regime, the $\rho_H$ does not significantly change  with the field, where a small, linear increase of $\rho_H$ is observed that  does not saturate in fields up to 9 T for both $H//ab$ plane and $H//c$-axis configurations. On the other hand, in the antiferromagnetic regime, {T = 200 K and 300 K,}  $\rho_H$ increases rapidly with the field {H$//c$ higher than the respective critical field for metamagnetic transition (Figs. 2e-2f)}. The overall behavior of Hall resistivity resembles the M(H) data shown in Fig. 2(c) and Fig. 2(d), respectively. 
 Furthermore, similarly to M(H) behavior, a narrow hysteresis of $\rho_H$ was observed in the AFM phase when the field is applied along with  $ab$ plane while no hysteretic behavior is observed for the field applied parallel to the $c$-axis (see Fig. S3 in SM \cite{SI} for more detail). \\
 
In topological magnetic materials, the total Hall resistivity ($\rho_{H}$) can be expressed as
\begin{flalign}
\noindent \rho_{H} = \rho^{N}+\rho^{A}+\rho^{T}, &&
\end{flalign}
where,  $\rho^{N}(H)  = R_0H$ is the normal Hall resistivity, { and R$_0$ is dubbed as the coefficient of normal Hall resistivity, which is defined  by the number of carriers (for a multiband metal system it is defined as the weighted average of the carriers with their mobilities)\cite{Ghimireeabe2680})}. Similarly,  $\rho^{A}(M) = R_{S}4\pi M$ is the anomalous Hall resistivity, and  $\rho^{T}$ refers to the topological Hall resistivity, which  originates from the Berry phase gained by electrons owing to spatially varying magnetization \cite{THE1, THE2}. 
Figure 3(a) shows the three different components of Hall resistivity at 200 K when $H//ab$. In the high field saturation region, Eq. 1 can be written as $\rho_H$ = R$_{0}$H + R$_S$4$\pi$M. The slope R$_{0}$ and intercept  4$\pi$R$_S$ can be estimated from the linear plot  of $\rho_H/M$ versus B/M in the high-field region. The perfect linear behavior of the $\rho_H/M$ versus B/M at high field indicates that the anomalous Hall resistivity ($\rho^{A}$) is the dominant component in the Hall resistivity. The R$_0$ derived for $H//ab$ is negative whereas it is positive for $H//c$ at 200 K which gives $\rho^{N}$ negative and positive, respectively as shown in Fig. 3(b). 
 {The sign change of R$_0$ could be the result of change in electronic structure as suggested in recent work \cite{PhysRevB.103.014416, wang2019near}, it requires further investigations.} The slopes 4$\pi$R$_S$ are found to be positive for both configurations.
 \begin{figure}
	\centering
	\includegraphics [width= 0.9\textwidth]{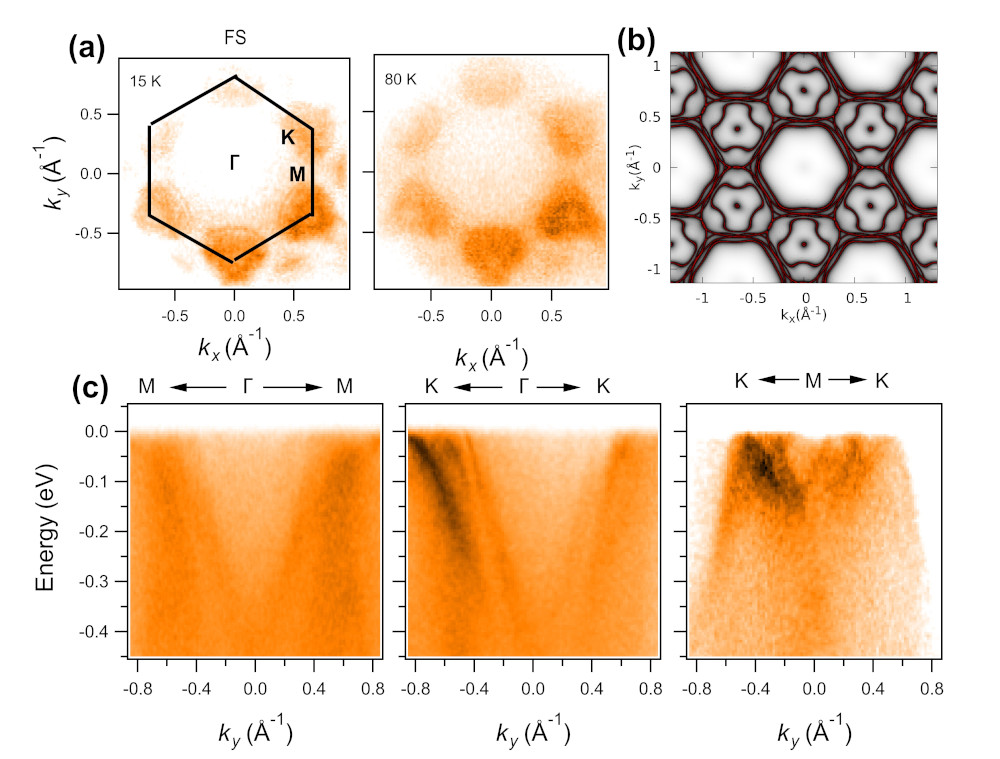}
	\caption{Electronic structure of ErMn$_{6}$Sn$_{6}$.
(\textbf{a}) {Fermi surface map  at different temperatures. The temperature values are noted in the plots.}    (\textbf{b}){ Calculated Fermi surface map.} (\textbf{c}){ Dispersion maps along the M-$\Gamma$-M, K-$\Gamma$-K, and K-M-K direction, respectively the FIM phase.}    ARPES measurements were performed at the SSRL beamline 5-2 using 100 eV photon energy with s-polarization. }
\end{figure}
The topological Hall resistivity ($\rho^{T}$)  is estimated by subtracting normal and anomalous components of the Hall resistivity from the total Hall resistivity and its contribution is small compared to anomalous Hall resistivity in both in-plane and c-axis configuration. The topological Hall effect is more pronounced when the magnetic field is applied in the $ab$-plane and it reaches to 1.8 $\mu\Omega$-cm at 1.5T (Fig. 3a, inset). The obtained value of $\rho^T$ along $ab$-plane in ErMn$_6$Sn$_6$ is 100\% higher and more importantly it observed at a much lower applied magnetic field compared to YMn$_6$Sn$_6$ compound reported in ref. \cite{ Ghimireeabe2680} (0.75 $\mu \Omega$-cm at H=4 T and T=200 K).  It is also much larger than reported for thin films correlated oxides \cite{RN125} and non-cubic AFM Mn$_5$Si$_3$ \cite{RN126}. Interestingly, a large anisotropic topological Hall effect of a similar magnitude to ErMn$_6$Sn$_6$ has been reported recently for the hexagonal non-collinear magnet, Fe$_5$Sn$_3$ \cite{doi:10.1063/5.0005493}. In addition, $\rho^T$ along $c$-plane exhibit significantly low value, but interestingly, it oscillates between negative 0.2 $\mu \Omega$-cm at 2.2 T to positive and $+$0.1 $\mu \Omega$-cm at 5 T (Fig. 3(b), inset). The critical magnetic field for peak in the  $\rho^T$ vs H  matches excellently with that obtained from M(H) at 200 K (see Fig. 2(c) inset and Fig. 2(d)). {The peak value of topological Hall resisitivity could be an indication of the presence of a new magnetic phase that may exhibit strong Hall resistivity. Further experimental and theoretical works are required to shed light on this intriguing result.} Furthermore,  the large difference of critical field between the $ab$ and $c$-plane  suggest the ErMn$_6$Sn$_6$ is highly anisotropy and which also play a crucial role in anomalous and topological Hall effect. 
The $\rho^{T}$ disappears in high field region (above saturation field) indicating that it cannot host a non-trivial spin structure. Interestingly, the signs of $\rho^T$ are opposite for the $H//ab$ and $H//c$ directions. The opposite signs of the topological Hall effect in different directions of the ErMn$_6$Sn$_6$ may be related to the anisotropic magnetic behavior in this material, which can induce opposite spin chirality along with the two directions \cite{doi:10.1063/5.0005493}.  
 \\

In order to investigate the electronic structures of ErMn$_{6}$Sn$_{6}$, we performed ARPES measurements using VUV-photon source. Figure 4(a) shows Fermi surface map parallel to the kagome plane, measured   at a temperature of 15 K of 80 K, respectively. The Fermi map shows the hexagonal symmetry as suggested by the crystal structure; the circular blobs are seen at the K points, manifesting an excellent hexagonal symmetry. The intense blobs probably appear as a consequence of manifold bulk bands. Several energy pockets at the edges of the hexagonal system indicate the complex band structure of this material, however all features are not discernible in the photoemission intensity plots presented in Fig. 4(a). 
We present the dispersion maps along the high symmetry directions i.e. M-$\Gamma$-M, K-$\Gamma$-K, and K-M-K directions, respectively. Electron like bands are seen along the  M-$\Gamma$-M and K-$\Gamma$-K directions.  The presence of an electron like pocket at the $\Gamma$ can be further confirmed by the constant energy contours presented in the SM \cite{SI}. Dispersion map along the M-$\Gamma$-M direction shows a quasi-two dimensional nature; as the band does not disperse with  photon energies (26-120 eV) \cite{SI}. The 2D-like bands can be attributed to be originated from the orbitals related to the kagome lattice (Mn) \cite{li2021dirac}. Our orbital-projected first principles calculations show that electron-like bands along the $\Gamma$-K direction are contributed by Mn  $d_{xz}$ and Mn $d_{z^2}$ (see Fig. S6 and Fig. S11 \cite{SI}).   The dispersion maps along the  K-M-K exhibit  the lower Dirac cone-like features at the K points.  The observation of Dirac cones at the K point, characteristics features  observed in many other kagome materials, suggests the possible  origin of the anomalous Hall effect in this material. Major ARPES features are captured in the first-principles calculations (see Fig. S11 for side by side comparison \cite{SI}), which {suggest} the presence of Dirac cone at the K point. { Presence of various factors such as  intrinsic magnetism, correlation, matrix elements effects, etc. play crucial roles for some discrepancies between ARPES and calculated plots.}   \\
 A 3D system with quasi-2D electronic structures can provide a condition for the anomalous Hall effect originated from  the Berry curvature.  First-principles calculations predict that the Chern gap lies above the Fermi level.  ARPES  measurements and first-principles calculations indicate the orbital-selective Dirac fermiology as seen in various kagome materials \cite{RN123, RN124, li2021dirac, yang2019evidence}. {As suggested by ref. \cite{clatterbuck1999magnetic}, the controlling interaction in ErMn$_6$Sn$_6$ is mainly contributed by Mn-Mn interactions, therefore one can anticipate the crucial role of the kagome layer in this system for exotic magnetic and transport properties.} Additionally, the flat bands cover almost all of the Brillouin zone, however they locate above the Fermi level. Thus, it does not contribute for the correlation. By possible electron-doping the flat band can be brought to Fermi level which could give rise to the correlated system. Further, replacing or doping with lighter compounds might add an interesting insight in these type of materials \cite{PhysRevB.103.144410}. 
In conclusion, by performing detailed magneto-transport measurements, we report a direct observation of anisotropically large anomalous and topological Hall effect in  ErMn$_{6}$Sn$_{6}$. Our combined experimental and theoretical investigations demonstrate  the multi-orbital kagome  band structure at the Fermi level.  Our study suggests a new platform to study the interplay of  correlation and magnetism in a kagome magnet system.
\\
\noindent \textbf{Acknowledgement}\\ 
M.N. is supported by the Air Force Office of Scientific
Research under award number FA9550-17-1-0415,
the Air Force Office of Scientific Research MURI (FA9550-20-1-0322), and the Center for Thermal Energy Transport under
Irradiation, an Energy Frontier Research Center funded
by the U. S. DOE, Office of Basic Energy Sciences. F.C.K and P.M.O. acknowledge support from the Swedish Research Council (VR) and the Knut and Alice Wallenberg Foundation (Grant No. 2015.0060). Computational resources were provided by the Swedish National Infrastructure for Computing (SNIC)(Grant No. 2018-05973).  Work at Buffalo State was supported by startup fund from SUNY Buffalo State College and the Office of Undergraduate Research Program. R.F. acknowledges financial support from Office of Undergraduate Research, EURO and Small Grant awards. N.P. and K.G. acknowledge support from the INL's LDRD program (18P37-008FP and 19P45-019FP, respectively) under DOE Idaho Operations Office Contract DE-AC07-05ID14517. We thank Donghui Lu, and Mokoto Hashimoto for beamline assistance.\\

$^*$These authors contributed equally in this work.\\
$^\dagger$ Corresponding author: Madhab.Neupane@ucf.edu

\newpage
\setcounter{figure}{0}
\renewcommand{\thefigure}{\textbf{S}\arabic{figure}}
\begin{quote}
	\centering
\textbf{Supplementary Materials}
\end{quote}
\bigskip
\bigskip

\bigskip
\noindent\textbf{1~~~~Methods}\\
\noindent\textbf{1.1~~~Crystal Growth}\\
\noindent The single crystals of ErMn$_6$Sn$_6$ were prepared by the flux growth technique with tin as the flux. The pure elements (99.9, 99.98, and 99.999 wt. for Er, Mn, and Sn, respectively; surface cleaned pieces) in a respective molar ratio of 1:6:30 were loaded in an alumina crucible. Then the crucible was sealed in a quartz tube under vacuum.  The quartz tube was heated to $1100^{\circ}$ C in 8 hrs, dwelled for 10 hrs, and then cooled down to $600^{\circ}$ C  over 140 hrs; finally centrifuged. Large flat hexagonal crystals, as large as 3-5 mm, were obtained. To verify the compositional homogeneity of crystals, we performed scanning electron microscopy (SEM) and energy dispersive X-ray (EDX) spectroscopy studies using a Leica Cambridge 360 microscope, equipped with an Oxford X-Max 20 EDX microprobe. The SEM and EDX results confirm the formation of a chemically homogeneous and stoichiometric 1:6:6 phase with no detectable impurities.
\begin{figure*}[h!]
\centering
	\includegraphics[width=0.78\textwidth]{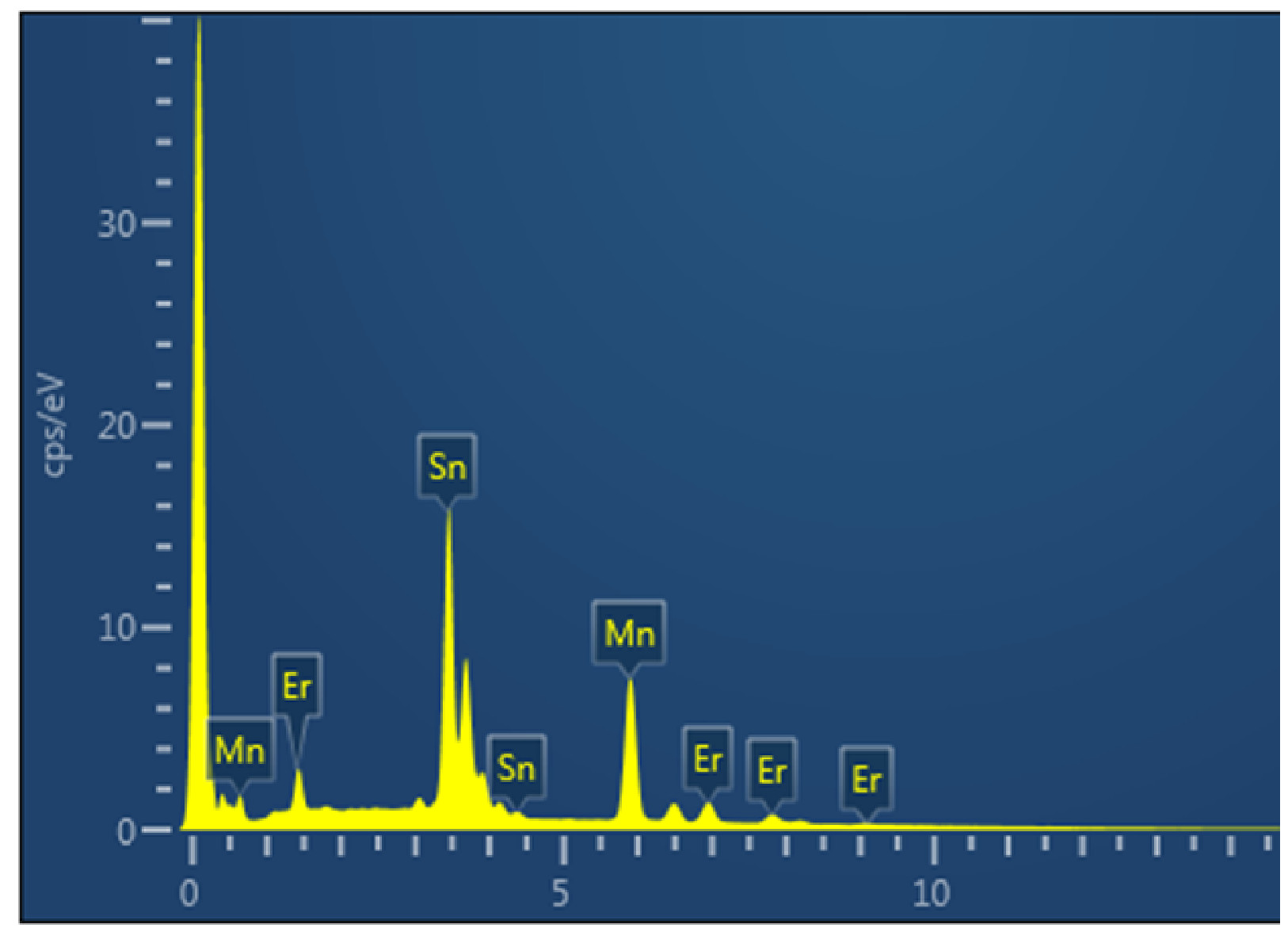}
	\begin{tabular}{|p{1.2cm}|p{1.5cm}|p{3.5cm}|p{1.5cm}|p{1cm}|p{1.5cm}|p{1.4cm}| } 
 \hline
 \multicolumn{7}{|c|}{Quant Results View}\\
 \multicolumn{7}{|c|}{viewed Data: Spectrum 2}\\
 \multicolumn{7}{|c|}{Processing Option Used: All Elements Processed}\\
 \hline
Element & Line Type & Apparent Concentration & k Ratio & Wt\% & Wt\% Sigma & Atomic\%\\
 \hline
 Mn & K series & 26.51 & 0.26508 & 27.42 & 0.29 & 46.00\\
 \hline
 Sn & L series & 56.21 & 0.56206 & 59.97 & 0.45 & 46.56\\
 \hline
 Er& L series & 11.22 & 0.11225 & 13.49 & 0.45 & 7.43\\
 \hline
 Total & & & & 100.88& & 100.00\\
 \hline
 \end{tabular}
	\caption{ Energy dispersive X-ray (EDX) spectroscopy analysis of ErMn$_6$Sn$_6$.
	}
\end{figure*}

\noindent\textbf{1.2~~~Magnetotransport measurement} \\ \noindent
Magnetic measurements of ErMn$_6$Sn$_6$ single crystals were carried out for both H$//$ab and H$//$c planes by Physical Property Measurement System (PPMS, Quantum Design), in the temperature range 2 to 400 K and magnetic field up to 9 T.  Electrical and Hall resistivity were measured by a standard four-probe technique with an electrical transport option available in PPMS.
\begin{figure*}[h!]
	\centering
	\includegraphics[width=0.8\textwidth]{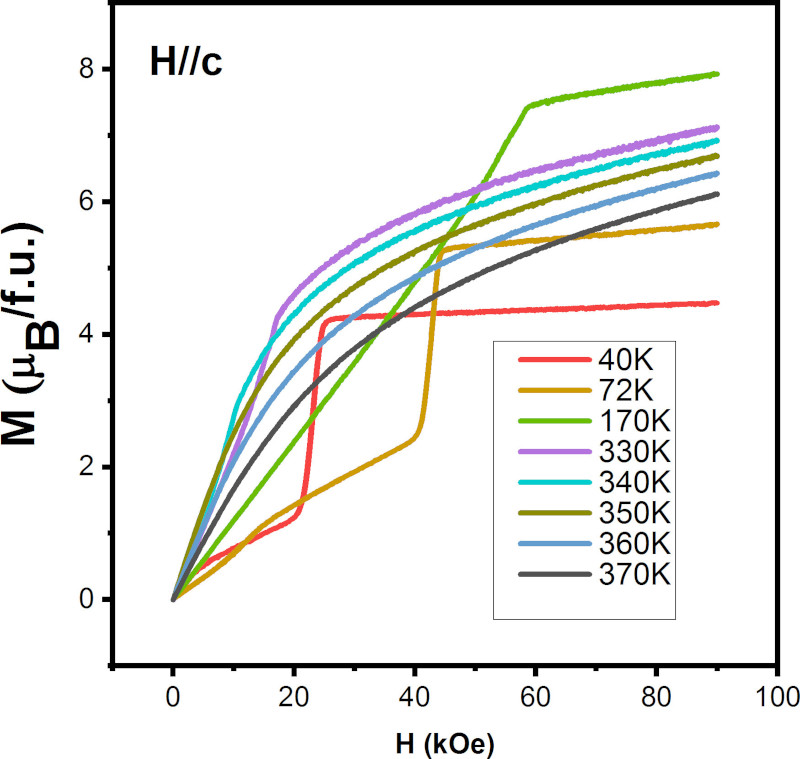}
	\caption{Magnetization M(H) as a function of magnetic field at different temperatures,
 measured along  H$//$c-axis.
	}
\end{figure*}
\begin{figure*}[h!]
	\centering
	\includegraphics[width=0.8\textwidth]{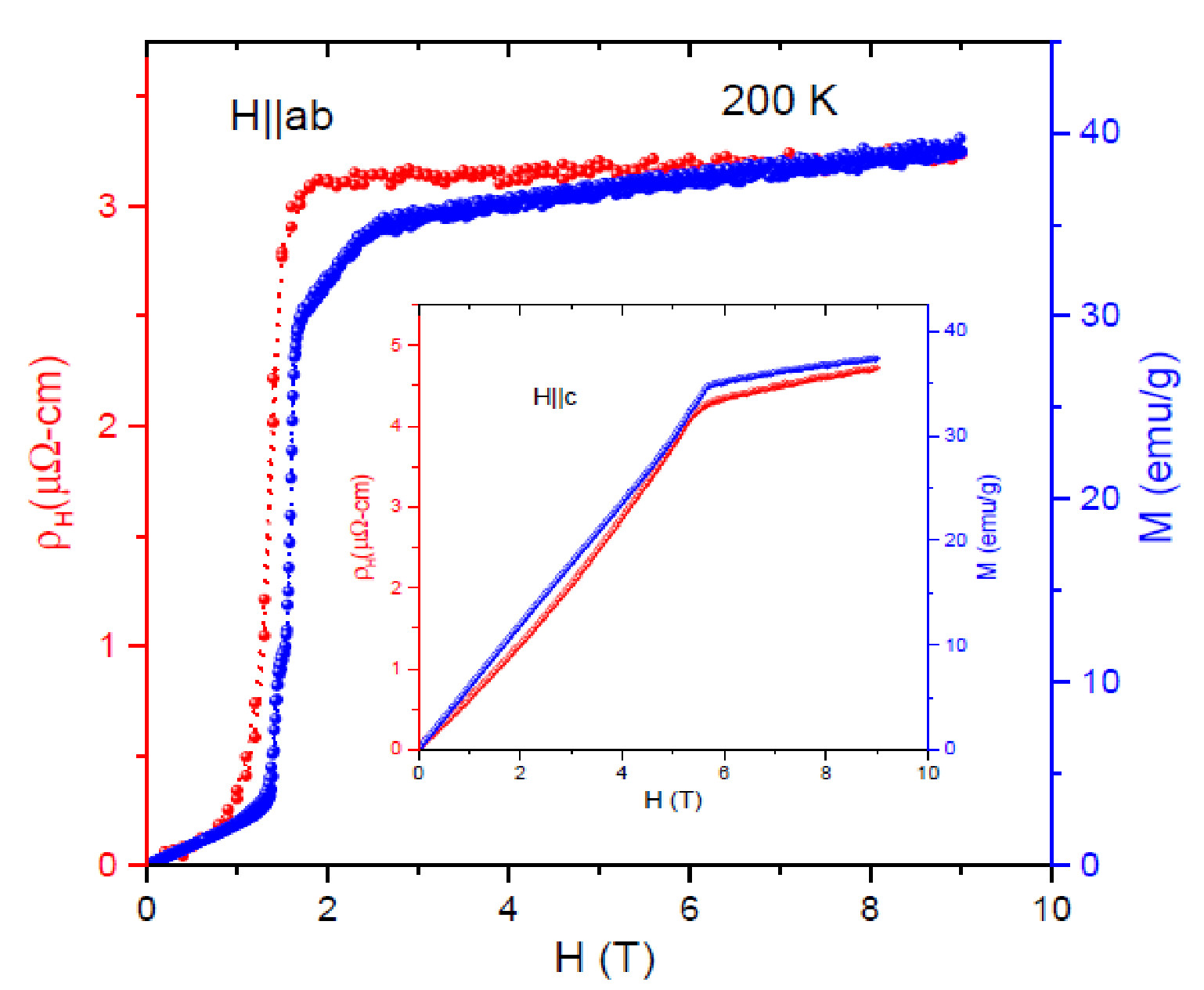}
	\caption{ Field dependence Hall resistivity ($\rho_{H}$ and magnetization M(H) for H$//$ab-plane and H$//$c-axis (inset) at 200 K.
	}
\end{figure*}

\noindent\textbf{1.3~~~ARPES measurement}\\
\noindent
Angle-resolved photoemission spectroscopy (ARPES) was performed at Stanford Synchrotron Radiation Lightsource (SSRL), Menlo park, CA. High quality flat and shining samples were used for measurements. The crystals were cut into small pieces and 
 mounted on
 copper posts. Torr seal was used to attach the samples to copper posts. Ceramic posts
were attached to the top of the samples. After,  samples were loaded into the chamber,
the chamber was cooled and pumped down for some hours. The crystals are cleaved in situ
at low temperature by knocking off the upper ceramic posts with a cleaver. Measurements
were carried out in temperatures in the range of 15 K-80 K. The pressure in the UHV was better than 1$\times$ 10$^{-10}$ torr. The angular and energy resolution was set better than 0.2$^{\circ}$ and 15 meV, respectively. Measurements were performed employing the photon energies in the range of 30 eV-150 eV using horizontal and vertical polarization. The measurements took more than 20 hrs without noticeable decay or damage of the samples.

\noindent\textbf{1.4~~~First-principles calculations}\\ \noindent
The first-principles electronic structure calculations were carried out within the density-functional formalism (DFT) \cite{Hohe64S,Kohn65S} as implemented in the Vienna ab initio simulation package (VASP) \cite{Kres96S,Kres96.1S}. The electron-ion interactions were modeled by the projector augmented wave potential \cite{Bloo94} and the exchange-correlation functional was approximated by the Perdew-Burke-Ernzerhof-type generalized gradient approximation (GGA) \cite{Perd96S}. The GGA + Hubbard U (GGA+U) approach  \cite{Duda98} with U = 8 eV on Er's 4f orbitals is used to treat with the exchange and correlation potential. The stability of the obtained results was checked against variation of the Hubbard U value; we found that  near the Fermi level, bands did not change for U in the range of 8 to 12 eV. Spin-orbit coupling is taken into account self-consistently. The cut-off energy of 500 eV and the 6$\times$6$\times$4 sampling of Brillouin Zone \cite{Monk76} are used and carefully checked to ensure the convergence. We have studied two different bulk configurations for ErMn$_6$Sn$_6$, including the ferrimagnetic (FIM) and anti-ferromagnetic (AFM) states. 
The surface states are calculated based on the Green's functions method \cite{Lope85}. Results from our DFT calculations were used as input to construct the tight-binding Hamiltonians from maximally localized Wannier functions (MLWF) using WAANIER90 \cite{Most14}. The tight-binding parametrs were calculated using an initial basis sets of Er (d), Mn (s, p, d) and Sn (s, p) orbitals, which resulted in an excellent fit of bulk band structure in the range of -2 eV to + 2 eV relative to the Fermi level E$_F$. The surface band structures were calculated using Wannier interpolation using the WANNIERTOOLS package \cite{Wu18S}.\\

\noindent\textbf{2~~~~ Hall resistivity and magnetic moment of ErMn$_6$Sn$_{6}$  }\\ \noindent
In Fig. S2, 2e present magnetization as a function of magnetic fields at different temperatures measured along the H$//$c-axis.   The field dependence of Hall resistivity and magnetization at 200 K is also shown in Fig. S 3. A large hysteresis in $\rho_{H}$ was observed in AFM phase at each temperature when the field is applied in ab-plane indicating the presence of in-plane ferromagnetic component. \\


\noindent\textbf{3~~~~ Calculated band structures }\\
\noindent\textbf{3.1~~~Bulk band structures in FIM and AFM}\\ \noindent
ErMn$_6$Sn$_{6}$ undergoes two different magnetic phases in the presence magnetic field, here we present calculations in  both phases ( see Figs. S4-S5) with  and without spin-orbit coupling (SOC). In the FIM state, linear band crossing are observed at K and M in the vicinity of the Fermi level, which are subjected to gap opening with the inclusion of the SOC. Bulk band calculations in the AFM state show the shifting of the bands without significant changes in the nature of the bands with the change of magnetic regimes. 
\begin{figure*}[h!]
	\centering
	\includegraphics[width=0.9\textwidth]{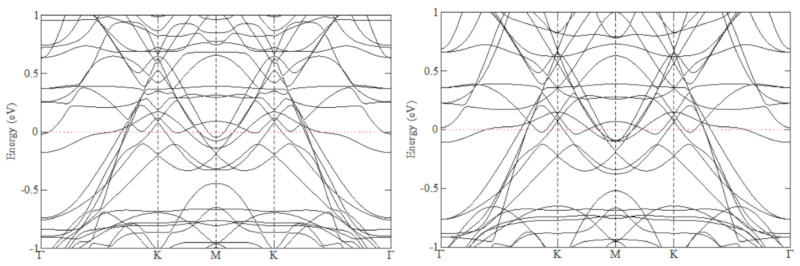}
	\caption{ Bulk band calculations in FIM magnetic phase (a) without SOC, and (b) with SOC, respectively.
	}
\end{figure*}
\begin{figure*}[h!]
	\centering
	\includegraphics[width=0.9\textwidth]{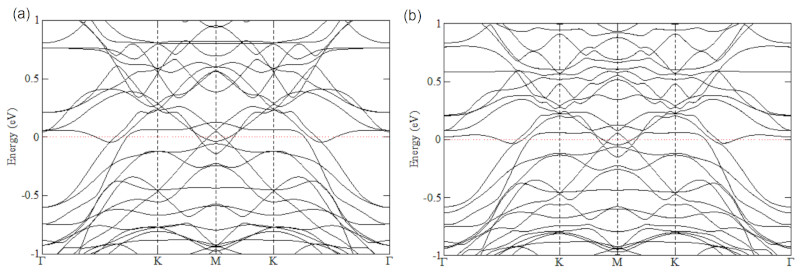}
	\caption{ Bulk band calculations in AFM magnetic phase (a) without SOC, and (b) with SOC, respectively.
	}
\end{figure*}

\noindent\textbf{3.2~~~Orbital nature of bulk band in FIM state}\\ \noindent
In Fig. S6, we present the orbital-projected bulk band structures of FIM  ErMn$_6$Sn$_{6}$, where we take account of \textit{d} bands of Mn. The calculations show the orbital origination of  kagome bands. Two electron-like bands along the K-$\Gamma$ are featured by $d_{xz}$ and $d_{z^2}$, which are denoted by blue and purple color, respectively. Similarly, Dirac-like bands above the Fermi level at the K point is formed by the $d_{xy}$ band, as represented by red color, whereas a Dirac-like dispersion below the Fermi level is featured by the $d_{z^2}$. Our calculations provide the evidence that the kagome system consists multi-orbitals in the vicinity of the Fermi level.
\begin{figure*}[h!]
	\centering
	\includegraphics[width=0.9\textwidth]{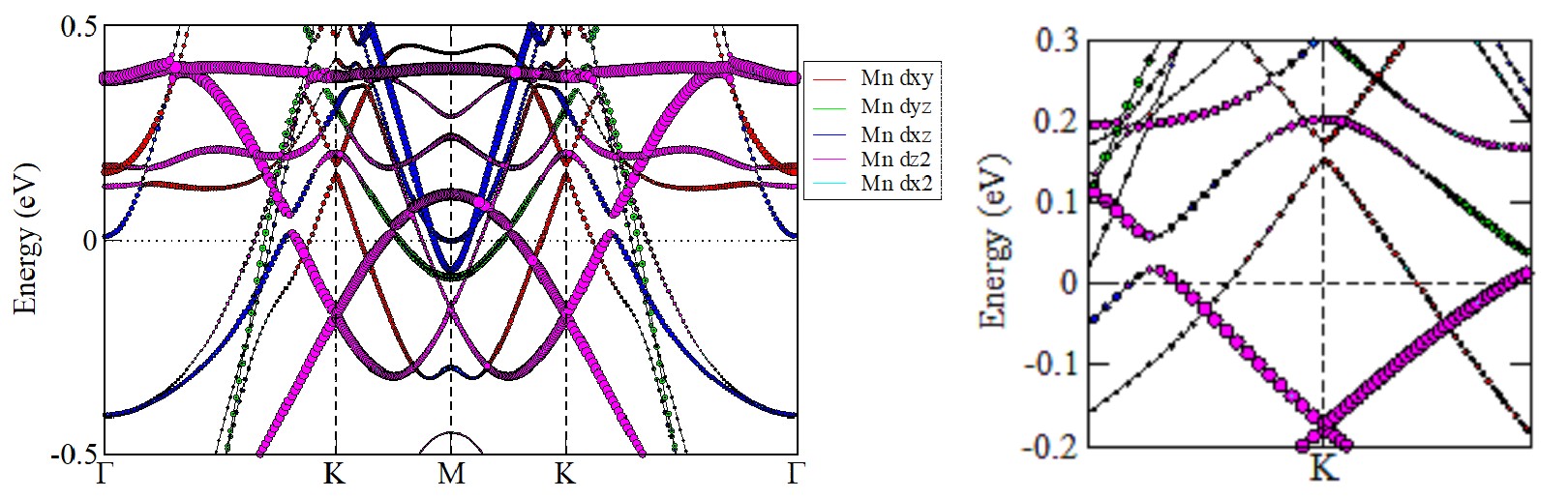}
	\caption{ Orbital-projected bulk band structure of FIM state along various high-symmetry directions. Right panel shows the zoom in view around Fermi level at K-point.
	}
\end{figure*}

\noindent\textbf{4~~~~Photoemission spectra of ErMn$_6$Sn$_6$, and surface calculations}\\ \noindent
Core-level  spectrum of the sample is presented in the Fig. S7, which shows the characteristics peaks for Sn 4\textit{d} and Mn 3\textit{p}. We present the Fermi map and constant energy contours of ErMn$_6$Sn$_6$ in Fig. S8, the Constant energy contours  show the complex metallic band structures. The constant energy show the hexagonal structures which are probably associated with the kagome crystal structures. The bands are electron-like at the $\Gamma$ point.  The shape of the Fermi surface  does not significantly change with the magnetic ordering show that surface bands are less affected by the magnetic ordering. However, the circular blobs change to almost triangular  at the K-point.\\

\begin{figure*}
	\centering
	\includegraphics[width=0.7\textwidth]{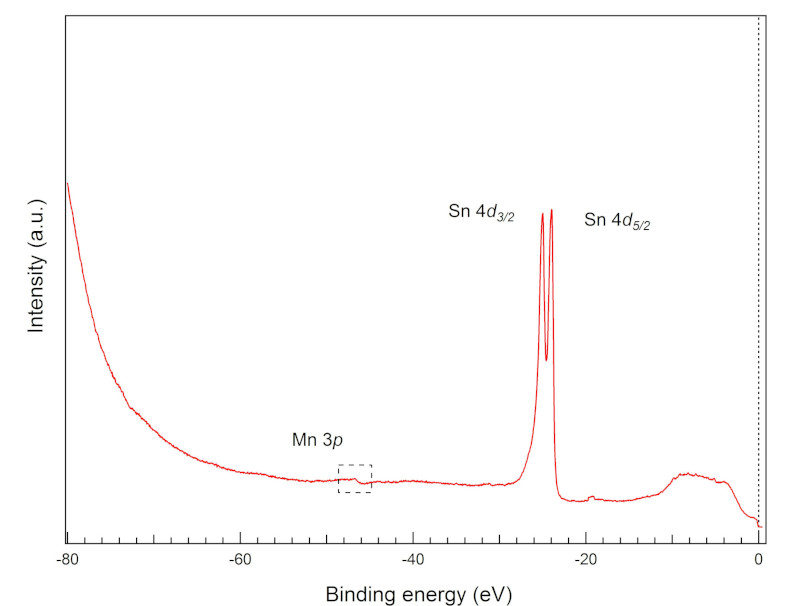}
	\caption{Core level spectrum of measured ErMn$_6$Sn$_6$ using photon energy of 100 eV.}
\end{figure*}

\begin{figure*}[h!]
	\centering
	\includegraphics[width=0.9\textwidth]{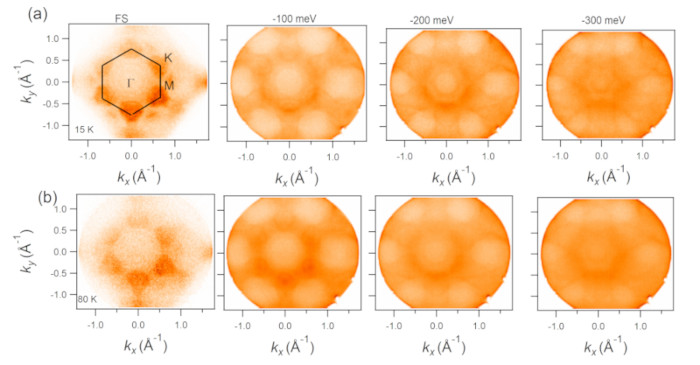}
	\caption{ Fermi map and constant energy contours measured using 100 eV photon energy at (a) low temperature (15 K), and (b) high temperature (80 K).
	}
\end{figure*}
\begin{figure*}[h!]
	\centering
	\includegraphics[width=0.8\textwidth]{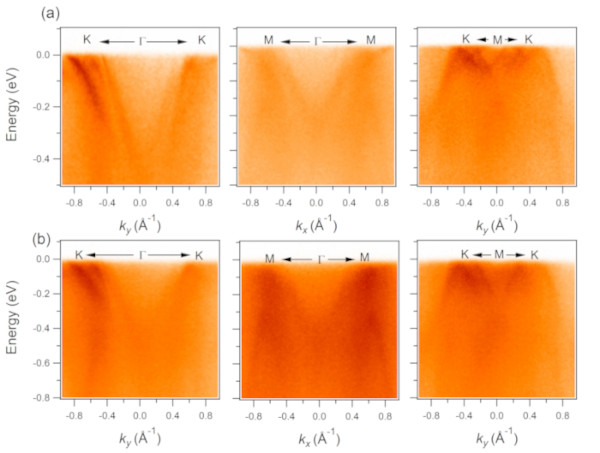}
	\caption{ Dispersion maps along the high symmetry directions at (a) low temperature (15 K), and (b) high temperature (80 K).
	}
\end{figure*}
\begin{figure*}[h!]
	\centering
	\includegraphics[width=0.8\textwidth]{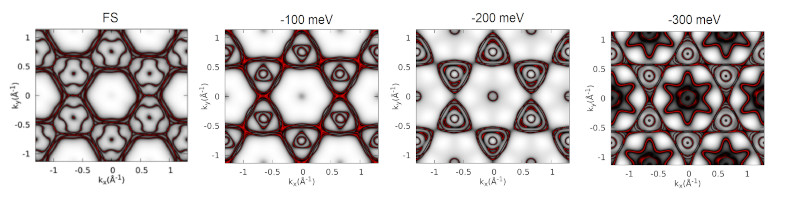}
	\caption{Calculated Fermi map and constant energy contours.
	}
\end{figure*}
\begin{figure*}[h!]
	\centering
	\includegraphics[width=0.9\textwidth]{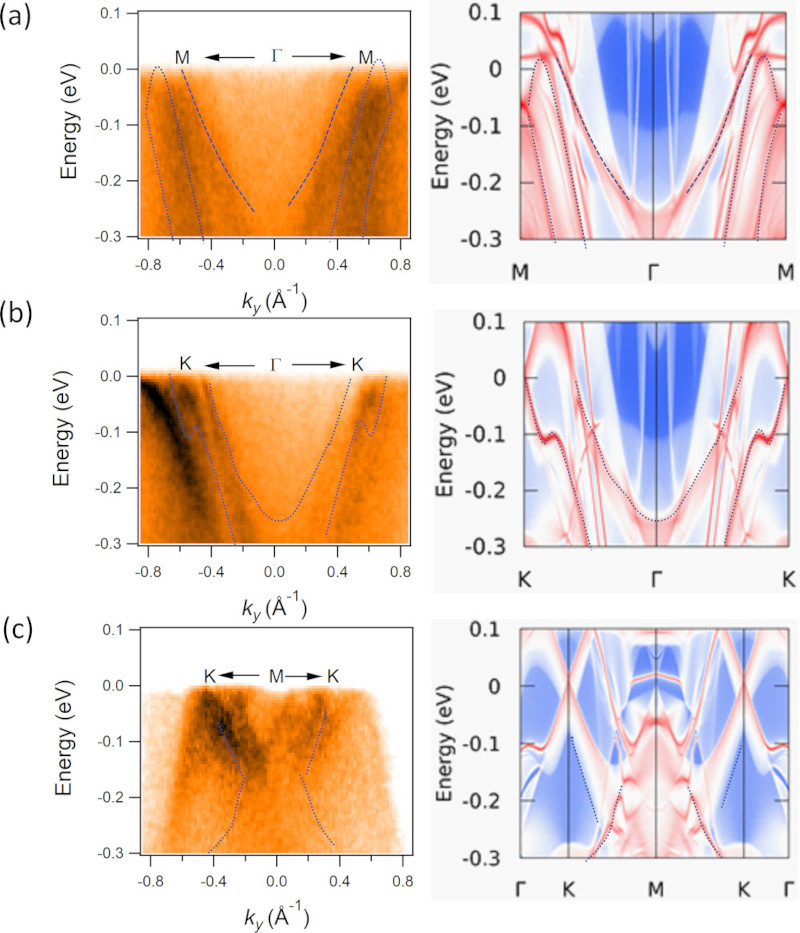}
	\caption{ Experimental and calculated dispersion maps along the (a) M-$\Gamma$-M, (b) K-$\Gamma$-K, and (c) K-M-K directions, respectively. Plot curve guide the eyes to compare experiments and calculations.  
	}
\end{figure*}
In Figure S10, calculated Fermi map and constant energy contours are presented which show the complex band structures associated with the kagome magnet.
 Calculations capture most of the features  seen on the measured maps, however, due to multiple-orbital nature of the bands, all bands cannot be seen on ARPES spectra. In ARPES, experimental geometry plays big roles, as a consequence some bands are suppressed or enhanced based on orbitals. We present experimental and calculated dispersion maps in Fig. S11, in which overlay plots serve as guide to the eyes. Our calculations capture  bands along the high symmetry directions. Furthermore, MDCs and EDCs provide the crucial information regarding the bands along the high-symmetry directions. We identify the parabolic-shaped bands  (electron-like bands) at the $\Gamma$ point which is  associated with the $d_{xz}$. 
\begin{figure*}[h!]
	\centering
	\includegraphics[width=0.9\textwidth]{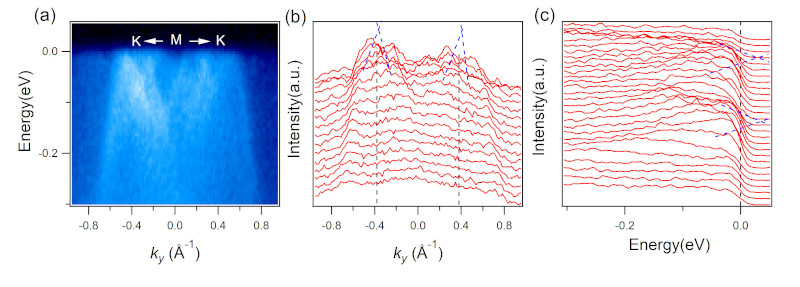}
	\caption{ (a) Dispersion map along the K-M-K direction. (b) MDCs along the K-M-K direction, the dotted line (black) refers to K-point. Blue dotted lines depict the projected Dirac cones at the K point. (c) EDCs along the K-M-K direction, dotted lines serve as guides to the eyes. Blue dotted lines depict the projected Dirac cones at the K point.
	}
\end{figure*}

\begin{figure*}[h!]
	\centering
	\includegraphics[width=0.8\textwidth]{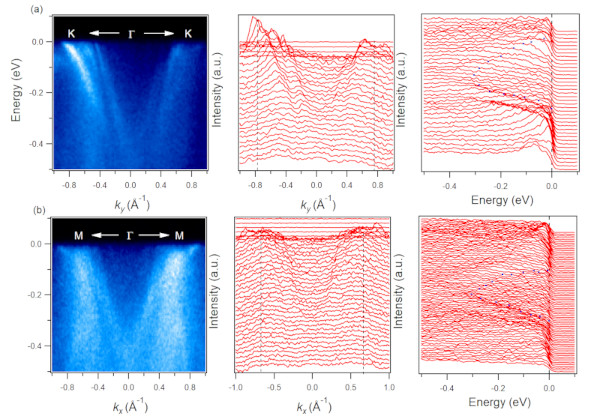}
	\caption{Dispersion maps and their MDCs and EDCs along the  (a) M-$\Gamma$-M, and (b) K-$\Gamma$-K direction respectively. Black dotted lines in the MDCs  show the location of M and K respectively. Similarly, the dotted curves in the EDCs work as the guides to the eyes to visualize the $d_{xz}$ as predicted by density functional theory.
	}
\end{figure*}

\noindent\textbf{5~~~$k_z$ dependent measurements}\\ \noindent
The photon energy dependent measurements provide the evidence of the surface states of a material. We use 26-120 eV photon energies to  explore the surface/bulk nature of the observed dispersion map along the $k_y$=0 direction which consists M-$\Gamma$-M direction. In Fig. S14, we present $k_z$-$k_x$ maps at two different photon energies. We use V$_{0}$ = 10 eV to calculate the out of the plane momentum. The parabolic band along the M-$\Gamma$-M does not significantly disperse with  photon energies suggesting quasi-two-dimensional nature. The photon energies used for measurements are VUV range, which provide extremely surface sensitive measurements.
\begin{figure*}
	\centering
	\includegraphics[width=0.9\textwidth]{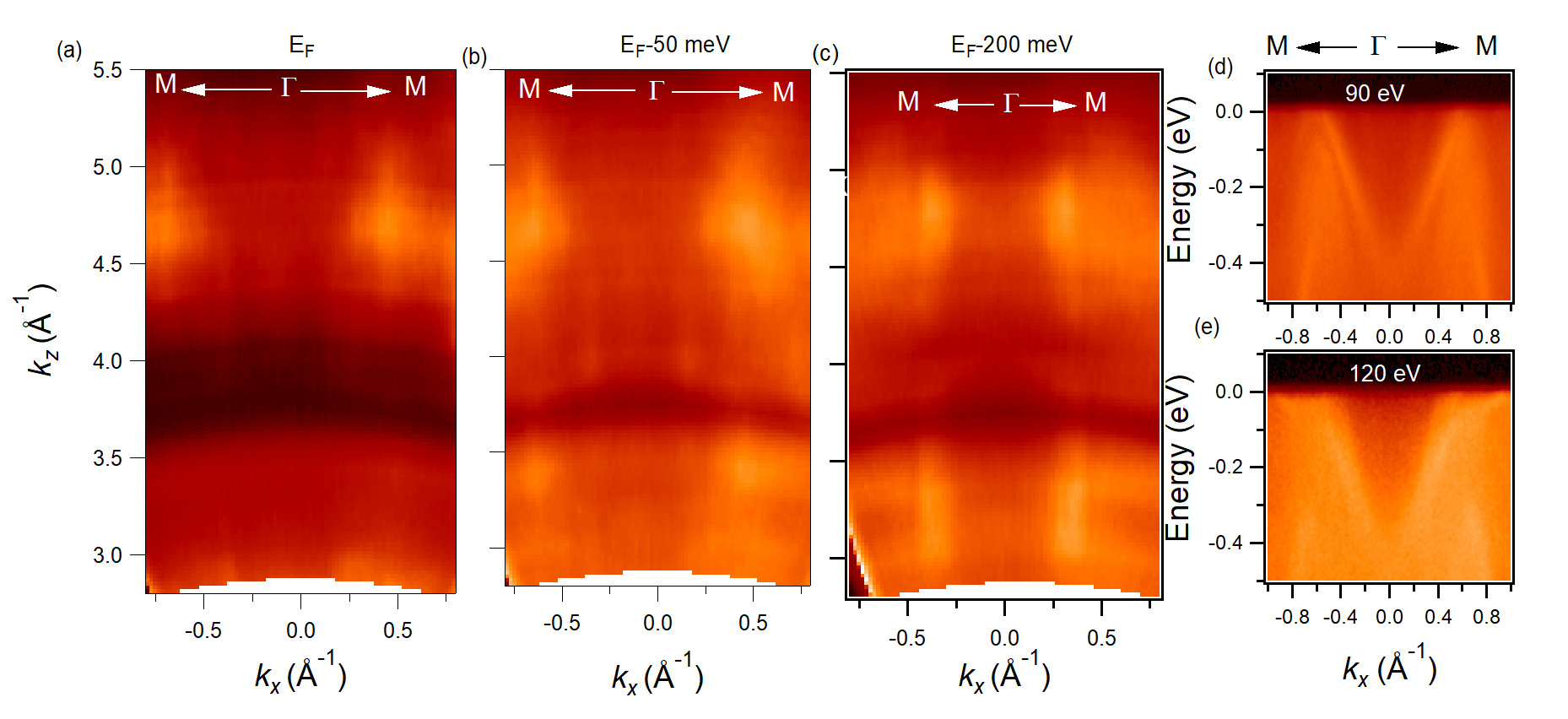}
	\caption{ $k_Z$ dependent measurement along $k_y$=0 (M-$\Gamma$-M) at the binding energy of (a) 0 eV,  (b) 50 meV, (c) 200 meV. Dispersion map along the M-$\Gamma$-M direction using photon energy of (d) 90 eV and (e) 120 eV. 
	}
\end{figure*}

\end{document}